\documentclass[aps,prl,showpacs,twocolumn,superscriptaddress]{revtex4}
\usepackage{graphicx}
\usepackage{bm}
\usepackage{amsmath}

\def\be{\begin{equation}}       \def\ee{\end{equation}}
\def\bea{\begin{eqnarray}}      \def\eea{\end{eqnarray}}

\begin{document}

\title{Semiclassical Time Evolution of the Holes from Luttinger Hamiltonian}
\author{Z. F. Jiang}
\affiliation{Beijing National Laboratory for Condensed Matter
Physics, Institute of Physics, Chinese Academy of Sciences,
Beijing 100080, China}
\author{R. D. Li}
\affiliation{School of Physics, Peking University, Beijing 100080,
China}
\author{Shou-Cheng Zhang}
\affiliation{Department of Physics, McCullough Building, Stanford
University, Stanford CA 94305-4045}
\author{W. M. Liu}
\affiliation{Beijing National Laboratory for Condensed Matter
Physics, Institute of Physics, Chinese Academy of Sciences,
Beijing 100080, China}

\begin{abstract}
We study the semi-classical motion of holes by exact numerical
solution of the Luttinger model. The trajectories obtained for the
heavy and light holes agree well with the higher order corrections
to the abelian and the non-abelian adiabatic theories in Ref.
\cite{sci} [S. Murakami \textit{et al.}, Science \textbf{301},
1378 (2003)], respectively. It is found that the hole trajectories
contain rapid oscillations reminiscent of the ``Zitterbewegung" of
relativistic electrons. We also comment on the non-conservation of
helicity of the light holes.
\end{abstract}
\maketitle

The field of spintronics holds the promise of using the spin
degree of freedom for building low-power integrated information
processing and storage devices \cite{pri,wolf}. Spintronics
devices also promises to access the intrinsic quantum regime of
transport, paving the path towards quantum computing. Recently, it
has been predicted theoretically that a dissipationless spin
current can be induced by an external DC electric field in a large
class of p-doped semi-conductors \cite{sci}. The dissipationless
spin current arises from the spin-orbit coupling in semiconductors
and several other groups have shown that it also applies to a
broader class of models \cite{c2,c1}.

The theory of Ref.\cite{sci} is based on the adiabatic solution to
the Luttinger model, which describes holes near the top of the
fourfold degenerate valence band. It was pointed out that the
abelian adiabatic approximation applies for the heavy-hole (HH),
while the non-abelian adiabatic approximation is required to
obtain the correct result for the light-hole (LH). The adiabatic
approximation is generally based on the separation of the light
and the heavy hole bands. However, at the top of the valence band,
these two bands intersect each other, and it is not clear to which
extent the adiabatic approximation is valid. In this paper, we
solve the semi-classical trajectory for the Luttinger model
exactly by numerical integration of the Heisenberg equation of
motion. We find that the full trajectory of the holes consists of
two parts, a rapidly oscillating part, reminiscent of the
``Zitterbewegung" of a relativistic electron \cite{zitter}, is
super-posed on a smooth part, which is accurately described by the
adiabatic theory. The separation of the rapid and the smooth parts
of the trajectory is also similar to the cyclotron and the guiding
center motion of a charged particle in an uniform magnetic field
and a spatially varying potential. In this sense, the adiabatic
approximation in the spin-orbit coupled systems is similar to the
lowest-Landau-level approximation in the quantum Hall effect.

The Luttinger effective Hamiltonian \cite{luttinger} with an d.c.
electric field $\mathbf{E} =E_{z} \widehat{z}$ can be written as
\cite{sci}
\begin{equation}
H=\frac{\hbar^{2}}{2m}((\gamma_{1}+\frac{5}{2}\gamma_{2})k^{2}-2\gamma
_{2}(\mathbf{k\cdot S})^{2})+eE_{z}z\newline. \label{1}%
\end{equation}
where $\gamma_{1}$, $\gamma_{2}$ are the valence-band parameters
for semiconductor materials. Luttinger \cite{luttinger} pointed
out that there are 16 linearly independent spin matrices which can
be chosen as $E$, $S_{x}$,
$S_{y}$, $S_{z}$, $S_{x}^{2}$, $S_{y}^{2}$, $\{S_{x},S_{y}\}$, $\{S_{y}%
,S_{z}\}$, $\{S_{z},S_{x}\}$, $\{S_{x},S_{y}^{2}-S_{z}^{2}\}$, $\{S_{y}%
,S_{z}^{2}-S_{x}^{2}\}$, $\{S_{z},S_{x}^{2}-S_{y}^{2}\}$,
$S_{x}^{3}$, $S_{y}^{3}$, $S_{z}^{3}$,
$S_{x}S_{y}S_{z}+S_{z}S_{y}S_{x}$. The full set of dynamic
variables in the theory consists of three position operators $x,y$
and $z$, three momentum operators $k_{x},k_{y}$ and $k_{z}$, and
the 16 spin matrices listed above. The Heisenberg equation of
motion for the expectation value of any operator $A$ is determined
by a differential equation $d\left\langle A\right\rangle /dt=
(i\hbar)^{-1}\left\langle [A,H]\right\rangle $. The equations of
motion for the momentum and the position operators are given by
\begin{equation}
\frac{d}{dt}\left[
\begin{array}
[c]{c}%
\left\langle k_{x}\right\rangle \\
\left\langle k_{y}\right\rangle \\
\left\langle k_{z}\right\rangle
\end{array}
\right]  =\frac{1}{\hbar}\left[
\begin{array}
[c]{c}%
0\\
0\\
c
\end{array}
\right]  , \label{2}%
\end{equation}
and%
\begin{align}
\frac{d}{dt}\left[
\begin{array}
[c]{c}%
\left\langle x\right\rangle \\
\left\langle y\right\rangle \\
\left\langle z\right\rangle
\end{array}
\right]   &  =\frac{2a}{\hbar}\left[
\begin{array}
[c]{c}%
\left\langle k_{x}\right\rangle \\
\left\langle k_{y}\right\rangle \\
\left\langle k_{z}\right\rangle
\end{array}
\right]  +\nonumber\\
&  \frac{b}{\hbar}\left[
\begin{array}
[c]{c}%
2\left\langle k_{x}S_{x}^{2}\right\rangle +\left\langle k_{y}\{S_{x}%
,S_{y}\}\right\rangle +\left\langle k_{z}\{S_{z},S_{x}\}\right\rangle \\
\left\langle k_{x}\{S_{x},S_{y}\}\right\rangle +2\left\langle k_{y}S_{y}%
^{2}\right\rangle +\left\langle k_{z}\{S_{y},S_{z}\}\right\rangle \\
\left\langle k_{x}\{S_{z},S_{x}\}\right\rangle +\left\langle k_{y}%
\{S_{y},S_{z}\}\right\rangle +2\left\langle
k_{z}S_{z}^{2}\right\rangle
\end{array}
\right]  , \label{3}%
\end{align}
where $a\equiv\hbar^{2}(\gamma_{1}+5\gamma_{2}/2)/2m, b\equiv-\hbar^{2}%
\gamma_{2}/m, c\equiv -eE_{z}$, and $\left\{  \text{ \ }\right\} $
represents the anticommutative relation. The equation of motion
for the spin operators can be obtained straightforwardly, but they
are lengthy and will not be given explicitly here.

Thus the evolution of momentum is simply determined by Eq.
(\ref{2}), and can be solved trivially analytically. Next we
numerically solve the equations for the spin operators, which
depends only on the solution of the momentum, not on the position.
Finally, we decompose the mean value of the product of the
momentum and the spin into the products of their mean values in
Eq. (\ref{3}), and numerically solve for the position operators.
For convenience, we can always choose a coordinate frame which
make the hole's initial momentum have no y-component.

\textbf{Time evolution of the heavy-hole:} The initial state of
the HH with helicity $\lambda=3/2$ can expressed as $\psi(0)=
U^{\dag}(\mathbf{k}(0))(
\begin{array}
[c]{cccc}%
1, & 0, & 0, & 0
\end{array}
)^{T}$, where $U^{\dag}(\mathbf{k}(0))=\exp(-i\phi
S_{z})\exp(-i\theta S_{y})$ is defined in Ref. \cite{sci}, and
$\mathbf{k}(0)$ is the initial momentum. The initial mean value of
any operator $A$ is $\left\langle A(0)\right\rangle
\equiv\left\langle \psi(0)\left\vert A\right\vert
\psi(0)\right\rangle $. From this definition of the initial state,
we obtain $\left\langle x(0)\right\rangle $, $\left\langle
k_{x}(0)\right\rangle $, $\left\langle S_{x}(0)\right\rangle $,
$\left\langle \{S_{x}(0),S_{y}(0)\}\right\rangle $ \textit{etc.}
as the initial conditions for Eqs. (\ref{2}, \ref{3}) and the spin
equations.

\begin{figure}[h]
\includegraphics[scale=0.7]{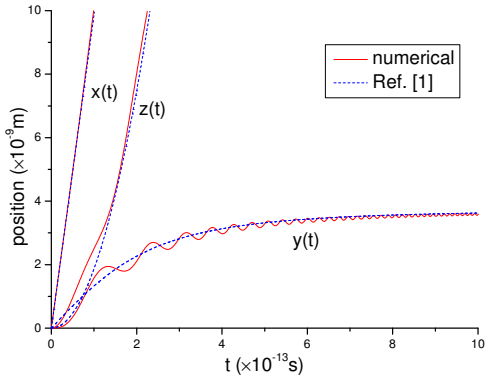}
\caption{A heavy-hole ($\lambda=3/2$) position three-component vs.
time. The red solid lines are numerical results, and the blue dash
lines are from the formulas of Ref. \cite{sci}. The initial
momentum is parallel to x-axis, and the electric field is parallel
to z-axis. These conditions are same in other figures.}
\label{f1}\end{figure}

In Fig. \ref{f1}, we plot the trajectory of the HH as a function
of time. We clearly see that besides the acceleration in the $z$
direction and the uniform velocity motion along the $x$ direction,
there is a side-way drift along the $y$ direction, which is
responsible for the dissipationless spin current. We can compare
the trajectories between our numerical solution and the result
from Ref. \cite{sci}. The abelian adiabatic equations of Ref.
\cite{sci} describe the overall trend very well. However, we see
that there are rapid oscillations on the exact numerical curve.
The frequency of the oscillations increases and the amplitude
decreases as the time increases. This ``Zitterbewegung" effect can
be obtained from the higher orders of the adiabatic approximation
theory \cite{sun90}. The oscillation on $z(t)$ can't be seen
clearly because the figure space is limited, but the oscillation
on $x(t)$ is really small which is analyzed as below.%

In order to study the higher orders of the adiabatic
approximation, we transform the Hamiltonian
(\ref{1}). We assume the wavefunction has the form of $\left\vert
\Psi(\mathbf{x},t)\right\rangle =\exp(-ieE_{z}zt/\hbar)\left\vert
u(\mathbf{k},t)\right\rangle $, then substitute $\left\vert \Psi(\mathbf{x}%
,t)\right\rangle $ into the Sch\"{o}dinger equation, so that we
get a new time-dependent Sch\"{o}dinger equation
$i\hbar\partial_{t}\left\vert
u(\mathbf{k},t)\right\rangle =H_{0}^{\prime}(t)\left\vert u(\mathbf{k}%
,t)\right\rangle $, where the new time-dependent effective
Hamiltonian
$H_{0}^{\prime}(t)=ak(t)^{2}+b(\mathbf{k}(t)\cdot \mathbf{S})^{2}$,where $\mathbf{k}%
(t)$ is determined by Eq.(\ref{2}). In the adiabatic
approximation, we assume
$\left\vert u(\mathbf{k},t)\right\rangle =\underset{\lambda}{\sum}C_{\lambda}%
(t)\exp(-\frac{i}{\hbar}\int_{0}^{t}\epsilon_{\lambda}(t^{^{\prime}%
})dt^{^{\prime}})U^{\dag}(\mathbf{k})\left\vert
\lambda\right\rangle $, where $\left\vert \lambda\right\rangle $
represents any eigenstate of $S_{z}$, so
$U^{\dag}(\mathbf{k})\left\vert \lambda\right\rangle $ is the
instant
eigenstate of $H_{0}^{\prime}(t)$. $H_{0}^{\prime}(t)U^{\dag}(\mathbf{k}%
)\left\vert \lambda\right\rangle =\epsilon_{\lambda}(t)U^{\dag}%
(\mathbf{k})\left\vert \lambda\right\rangle $, where
$\epsilon_{\lambda }(t)=\frac{\hbar^{2}k(t)^{2}}{2m_{\lambda}}$.
We substitute $\left\vert u(\mathbf{k},t)\right\rangle $ into the
time-dependent Sch\"{o}dinger equation, so that we get the
equation of $C_{\lambda}(t)$ is
\begin{equation}
\frac{d}{dt}C(t)+B\cdot C(t)=0, \label{im}%
\end{equation}
where $C(t)\equiv\left(
\begin{array}
[c]{cccc}%
C_{\frac{3}{2}} & C_{\frac{1}{2}} & C_{-\frac{1}{2}} & C_{-\frac{3}{2}}%
\end{array}
\right)  ^{T}$, and
\begin{equation}
B\equiv\left(
\begin{array}
[c]{cccc}%
0 & -\frac{\sqrt{3}}{2}\overset{\cdot}{\theta}e^{-i\alpha} & 0 & 0\\
\frac{\sqrt{3}}{2}\overset{\cdot}{\theta}e^{i\alpha} & 0 &
-\overset{\cdot
}{\theta} & 0\\
0 & \overset{\cdot}{\theta} & 0 &
-\frac{\sqrt{3}}{2}\overset{\cdot}{\theta
}e^{i\alpha}\\
0 & 0 & \frac{\sqrt{3}}{2}\overset{\cdot}{\theta}e^{-i\alpha} & 0
\end{array}
\right)  ,
\end{equation}
where
$\alpha\equiv\frac{1}{\hbar}\int_{0}^{t}\Delta\epsilon(t^{\prime
})dt^{\prime}$ is the dynamic phase, and $\Delta\epsilon(t^{\prime}%
)\equiv\epsilon_{L}(t^{\prime})-\epsilon_{H}(t^{\prime})$ is the
energy difference of HH and LH. If we choose the initial state
$C(0)\equiv\left(
\begin{array}
[c]{cccc}%
1, & 0, & 0, & 0
\end{array}
\right)  ^{T}$ , the adiabatic approximation assumes that
$0\approx C_{-\frac{3}{2},\pm\frac{1}{2}}(t)\ll
C_{\frac{3}{2}}(t)\approx1$ is always
satisfied. So only one equation remains, $\frac{d}{dt}C_{\frac{1}{2}}%
(t)=\frac{\sqrt{3}}{2}\overset{\cdot}{\theta}e^{i\alpha}$. We can
solve it
after the approximation that both $\Delta\epsilon$ and $\overset{\cdot}%
{\theta}$ are slowly varying functions of $t$. Then the
first-order correction of trajectory is\begin{figure}[h]
\includegraphics[scale=0.7]{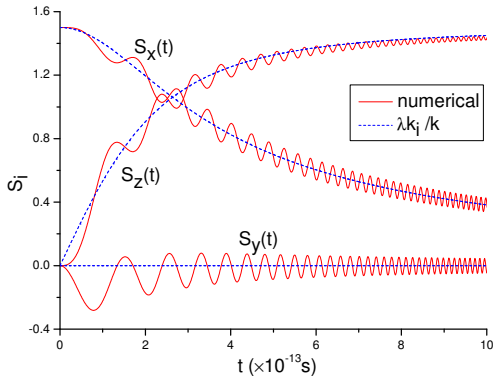}
\caption{The mean values of spin three-component of a heavy-hole
($\lambda=3/2$) vs. time. The red solid lines are numerical
results, and the blue dash lines are $\lambda k_{i}/k.$}
\label{f2}\end{figure}
$\mathbf{x}^{(1)}=C_{\frac{3}{2}}^{\ast}(t)C_{\frac{1}{2}%
}(t)\cdot e^{-i\alpha}\cdot\left\langle \frac{3}{2}\right\vert U(\mathbf{k}%
)i\frac{\partial}{\partial\mathbf{k}}U^{\dag}(\mathbf{k})\left\vert
\frac {1}{2}\right\rangle +h.c.$. This method is applicable to the
other three kinds of holes, too. So we get the unified formulas of
the first-order correction on the trajectory of any helicity state,%

\begin{align}
x^{(1)}  &  =\frac{\lambda(2\lambda^{2}-\frac{7}{2})eE_{z}\sin2\theta}%
{2k^{2}\Delta\epsilon}(1-\cos(\frac{\Delta\epsilon}{\hbar}t)),\nonumber\\
y^{(1)}  &  =-\frac{\lambda(2\lambda^{2}-\frac{7}{2})eE_{z}\sin\theta}%
{k^{2}\Delta\epsilon}\sin(\frac{\Delta\epsilon}{\hbar}t),\label{osci}\\
z^{(1)}  &
=\frac{\lambda(2\lambda^{2}-\frac{7}{2})eE_{z}\sin^{2}\theta
}{k^{2}\Delta\epsilon}(1-\cos(\frac{\Delta\epsilon}{\hbar}t)),\nonumber
\end{align}

From these formulas, we can see that the frequency $\omega=$
$\Delta \epsilon/\hbar$ will increase while the amplitude
$(k^{2}\Delta\epsilon)^{-1}$ will decrease as time increases, as
shown in Fig. \ref{f1}. We can evaluate the quantities of
frequency and amplitude in Fig. \ref{f1}, which agree with Eq.
(\ref{osci}) very well. The oscillation on $x(t)$ is small because
$\sin2\theta\approx0$.

Now let's study the applicability of Eq. (\ref{osci}). We have
used the approximation that both $\Delta\epsilon$ and
$\overset{\cdot}{\theta}$ are slowly varying functions of $t$,
which is equivalent to $\overset{\cdot
}{\Delta\epsilon}dt\ll\Delta\epsilon$ and $\overset{\cdot\cdot}{\theta}%
dt\ll\overset{\cdot}{\theta}$. They imply the same result,
$eE_{z}\Delta t\ll\hbar k$, which means that the approximation is
valid when the electric
field has not brought large changes in momentum. If we assume $E_{z}%
=1\times10^{3}V/m$, and $k=4\times10^{8}m^{-1}$, we get $\Delta
t\ll2000T$. So we have enough periods of oscillations in which Eq.
(\ref{osci}) is applicable.

\begin{figure}[h]
\includegraphics[scale=0.4]{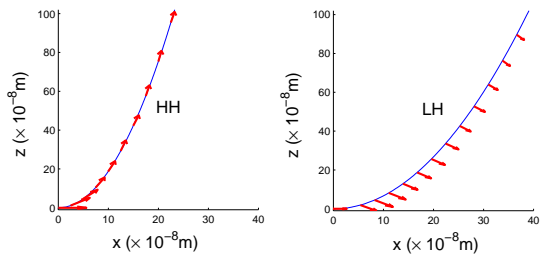}
\caption{Comparison of the trajectories and spins between
heavy-hole and light-hole.}
\label{m2}\end{figure}

Fig. \ref{f2} indicates $S_{i}(t)\approx\lambda k_{i}(t)/k(t)$,
which implies the approximate conservation of HH's helicity. This
can be seen clearly in Fig. (\ref{m2}). The oscillations show that
the semiclassical spin vector always precesses around the momentum
direction as the momentum changes in an electric field. The
oscillation can be calculated with the similar method above. The
deep reason for the HH's helicity conserving is the matrix element
representing transition between $\lambda=\pm3/2$ is zero. But the
LH's helicity isn't conserved as shown in the next section.

\begin{figure}[h]
\includegraphics[scale=0.7]{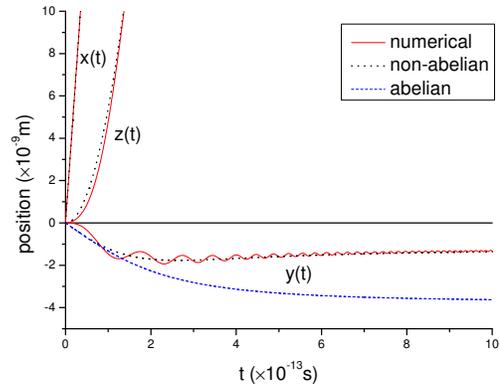}
\caption{A light-hole ($\lambda=1/2$) position three-component vs.
time. The red solid lines are numerical results, the blue dash
line is from the abelian adiabatic theory, and the black dot lines
are from the non-abelian adiabatic theory.}
\label{f3}\end{figure}

\begin{figure}[h]
\includegraphics[scale=0.7]{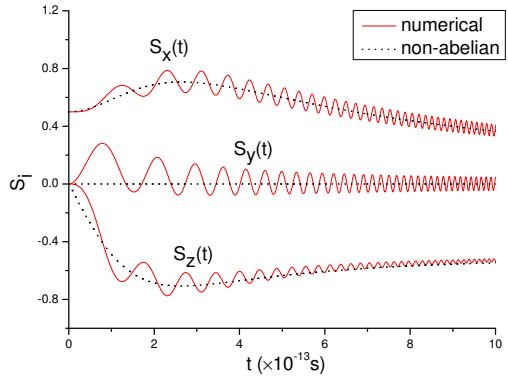}
\caption{The mean values of spin three-component of a light-hole
($\lambda=1/2$)\ vs. time. The red solid lines are numerical
results, and the black dot lines are from the non-abelian
adiabatic theory.}
\label{f4}\end{figure}

\textbf{Time evolution of the light-hole:} When we choose the
initial state as
$\psi(0)=U(\mathbf{k}(0))^{\dag}(%
\begin{array}
[c]{cccc}%
0, & 1, & 0, & 0
\end{array}
)^{T}$, Eqs. (\ref{2}, \ref{3}) and the spins' equations describe
the evolution of a LH with helicity $\lambda=1/2$. The trajectory
is shown in Fig. \ref{f3}, and the evolution of spin is showed in
Fig. \ref{f4}. The anomalous shift in y-direction is not as large
as predicted from the abelian adiabatic theory of Ref. \cite{sci},
and the helicity is no longer as conserved as that of HH. However,
both the trajectory and the evolution of spin can be explained in
the non-abelian adiabatic theory \cite{sci,sun90}, which properly
takes
into account the transition between the two LH states.%

If we confine the problem in the light hole's space, Eq.
(\ref{im}) is reduced to
\begin{equation}
\frac{d}{dt}\left(
\begin{array}
[c]{c}%
C_{\frac{1}{2}}\\
C_{-\frac{1}{2}}%
\end{array}
\right)  +\left(
\begin{array}
[c]{cc}%
0 & -\overset{\cdot}{\theta}\\
\overset{\cdot}{\theta} & 0
\end{array}
\right)  \left(
\begin{array}
[c]{c}%
C_{\frac{1}{2}}\\
C_{-\frac{1}{2}}%
\end{array}
\right)  =0.
\end{equation}
It describes the evolution of two degenerate states. The solution
is
\begin{equation}
C(t)=\left[
\begin{array}
[c]{cc}%
\cos(\theta_{t}-\theta_{0}) & \sin(\theta_{t}-\theta_{0})\\
-\sin(\theta_{t}-\theta_{0}) & \cos(\theta_{t}-\theta_{0})
\end{array}
\right]  C(0), \label{solution}%
\end{equation}
where $\theta_{t}$ is the the polar angle at the time $t$. So we
can get the
anomalous shift in y directions%
\begin{align}
y_{\pm\frac{1}{2}}(t)  &  =C^{\dagger}(t)U(\mathbf{k})\cdot i\partial_{k_{y}%
}U^{\dagger}(\mathbf{k})C(t)\nonumber\\
&  =\pm\frac{3\cos(\theta_{t}-2\theta_{0})-\cos(3\theta_{t}-2\theta_{0}%
)-2\cos\theta_{0}}{4k_{0}\sin\theta_{0}}, \label{y}%
\end{align}
and the evolution of spin is $\left\langle
\mathbf{S}(t)\right\rangle
=C^{\dagger}(t)U(\mathbf{k})\mathbf{S}U^{\dagger}(\mathbf{k})C(t)$,
\begin{align}
S_{x,\pm\frac{1}{2}}(t)  &
=\mp\lbrack\frac{3}{4}\sin(\theta_{t}-2\theta
_{0})+\frac{1}{4}\sin(3\theta_{t}-2\theta_{0})],\nonumber\\
S_{y,\pm\frac{1}{2}}(t)  &  =0,\label{spin}\\
S_{z,\pm\frac{1}{2}}(t)  &
=\pm\lbrack\frac{3}{4}\cos(\theta_{t}-2\theta
_{0})-\frac{1}{4}\cos(3\theta_{t}-2\theta_{0})].\nonumber
\end{align}

The results from Eq. (\ref{y}) and (\ref{spin}) has been plotted
in Fig. \ref{f3} and Fig. \ref{f4}, they describe the trends of
numerical curves very well except for the rapid oscillation on the
numerical curves, which have been explained in the previous
sections due to higher order corrections to the adiabatic theory.

At last, we obtain the anomalous velocity in y-direction,
\begin{equation}
v_{y,\pm\frac{1}{2}}(t)= \pm\frac{3eE_{z}}{4\hbar k^{2}}
[\sin(\theta
_{t}-2\theta_{0}) -\sin(3\theta_{t}-2\theta_{0})]. \label{vy}%
\end{equation}
When $t=0$, $v_{y,\pm\frac{1}{2}}(0)=\lambda(2\lambda^{2}-\frac{7}{2}%
)eE_{z}k_{x0}/(\hbar k_{0}^{3})$, which is just the Eq. (7) of
Ref. \cite{sci} (where $F_{ij}$ is given by Eq. (S5) of SOM). Eq.
(\ref{vy}) represent the anomalous velocity at any time.

Unlike the HH, the LH does not always stay as an eigenstate, it
will evolute according to Eq. (\ref{solution}). Fig. (\ref{m2})
compares the spins' evolution of HH and LH. Obviously, LH's
helicity is not as conserved as HH, so LH's spin can't be always
parallel to its momentum like HH. The non-abelian adiabatic theory
of Ref. \cite{sci} properly takes this effect into account.

\textbf{The adiabatic condition:} Ref. \cite{debate} raised a
criticism by asking why the anomalous shift in Ref. \cite{sci} is
independent of $\lambda_{2}$. Actually, if $\lambda_{2}=0$, the
anomalous shift vanishes because the Hamiltonian degenerates to an
ordinary one without the spin-orbit coupling, and the adiabatic
approximation is no longer valid. Below can we see explicitly that
the adiabatic approximation fails when $\lambda_{2}$ is less than
a certain quantity. The condition of adiabatic approximation
\cite{sun90} is
\begin{equation}
\left\vert \frac{\left\langle H,\alpha\left\vert
\frac{d}{dt}\right\vert L,\beta\right\rangle
}{\frac{E_{H}-E_{L}}{\hbar}}\right\vert =\frac
{\frac{\sqrt{3}}{2}\frac{d\theta}{dt}}{2k^{2}\frac{b}{\hbar}}=\frac{\sqrt
{3}meE_{z}\sin\theta}{4k^{3}\hbar^{2}\gamma_{2}}\ll1. \label{condition}%
\end{equation}

The condition is better satisfied if $E_{z}$ is smaller and
$\gamma_{2}$ is larger. \ The small $E_{z}$ ensures that the
time-dependent Hamiltonian changes slowly, and the large
$\gamma_{2}$ ensures that the energy difference
between the HH and LH bands is large ($\Delta\epsilon=\frac{2\hbar^{2}%
k^{2}\gamma_{2}}{m}$), so the transition probability between HH
and LH is small.

In most semiconductors, Eq. (\ref{condition}) can be satisfied.
For example as GaAs, $\gamma_{2}=1.01$,
$k_{F}\approx8\times10^{8}m^{-1}$, if we assume $E_{z}=10^{3}V/m$,
$\theta_{0}=90^{o}$, we get the condition is $k\gg 0.02k_{F}$. So
only a little part in the middle of Fermi Ball doesn't meet the
conditions. We can neglect them when we integrate the whole Fermi
ball.

\textbf{Conclusion} We have studied the motions of the heavy-hole
and light-hole in a large class of hole-doped semiconductor based
on the Luttinger Hamiltonian. The trajectory of HH has rapid,
small amplitude oscillations, which can be explained as the
first-order correction on the trend described in Ref. \cite{sci}.
The trajectory of LH is more complicated and the helicity of the
LH is not conserved. The non-conservation of the helicity
invalidates the abelian adiabatic approximation. However, the
motion of LH can be well explained by the non-abelian adiabatic
theory. The excellent agreement between the exact numerical
solution of the Heisenberg equation of motion and the adiabatic
approximation validates the key assumptions leading to the
dissipationless spin current, and addresses the naive criticism
raised in Ref. \cite{debate}. In the future, we plan to apply the
formalisms developed in this paper to study the Luttinger
Hamiltonian under more general external conditions.

We thank Andrei Bernevig, D. Culcer and Q. Niu for helpful
discussion. This work was supported by the NSF of China under
grant number 10174095 and 90103024, the US NSF under grant number
DMR-0342832, and the US Department of Energy, Office of Basic
Energy Sciences under contract DE-AC03-76SF00515.

\end{document}